\documentclass[pre,aps,epsfig,twocolumn,showpacs]{revtex4}
\usepackage{graphicx}
\begin{document}

\title{Kinetics of the $A+B \to 0$ reaction with mass-dependent fragmentation}
\author{Sungchul Kwon and Yup Kim}
\affiliation{Department of Physics and Research Institute of Basic
Sciences, Kyung Hee University, Seoul 130-701, Korea}

\date{\today}

\begin{abstract}
We investigate the kinetics of uniformly driven
$A+B \to 0$ reaction with mass-dependent fragmentation in one dimension.
In this model, the fragmented mass $m$ of a site with mass $n_i$ is given
as $m=n^{\lambda} _i$, and it is driven to the one direction. When opposite species masses
occupy the same site, mass reaction takes place instantaneously.
Since the fragmented mass $m$ of $\lambda<1$ is less than mass $n_i$ of a site, the exponent
$\lambda$ controls the attractive interaction between particles at the same site.
The $\lambda=0$ case corresponds to hard-core (HC) particle system.
With equal initial densities of both species, we numerically confirm that
the scaling behaviors of density and lengths except the domain length $\ell$ are the
same as that of the uniformly driven HC particle system.
The scaling behavior of $\ell$ is $\ell \sim t^{2/3}$.
The kinetics of the reaction is independent of $\lambda$ as long as $\lambda<1$.
The $\lambda$-independent kinetics results from the $\lambda$-independent collective
motions of single species domains.

\end{abstract}

\pacs{05.70.Ln,05.40.-a} \maketitle

\section{Introduction}
The irreversible two species reaction $A+B \to 0$ has been widely
and intensively investigated among diffusion-limited reactions
because of its rich kinetics and wide applications to various
phenomena such as physics, chemistry and biology
\cite{chem,phy,privman,first,monopole,kang}.
The reaction instantaneously takes place with a rate $k$ when two
particles of opposite species encounter on the same site. For
homogeneous initial distributions with equal densities of $A$ and
$B$, $\rho_A(0) = \rho_B (0)$, the density $\rho (t)$ algebraically
decays in time $t$. In higher dimensions than the upper critical
dimensions ($d \geq d_c$), the kinetics follows the mean-field rate
equation and $\rho$ decays as $\rho \sim t^{-1}$ in time $t$. However for $d <
d_C$, the kinetics is fluctuation-dominated and depends on the
motion and the mutual statistics of particles
\cite{first,monopole,kang,length,namb,rg,kwon06,HC1,HC2,HC3,HC4,jwlee,rgabhc,AB-RRC}.

For isotropic diffusions, $\rho(t)$ scales as $\rho (t)
\sim t^{-d/4}$ with $d_c = 4$ \cite{monopole,kang,length,namb,rg}.
However, when the diffusion of two species is anisotropic, the
kinetics is completely changed. With the global relative drift of
one species to the other, $\rho (t)$ scales as $\rho (t) \sim t^{-(d+1)/4}$ for
$d \leq 3$ \cite{kang}. When only boundary particles of domains are
relatively driven, the kinetics
continuously changes according to the form of the drift velocity
\cite{kwon06}. For all cases mentioned above, the hard-core (HC)
constraint between identical particles is irrelevant to the
asymptotic kinetics. However, when both species are uniformly driven
to the same direction, the HC constraint completely changes the
kinetics. In one dimension, Janowsky reported that $\rho$ scales as
$t^{-1/3}$ for HC particles instead of $t^{-1/4}$ \cite{HC1}. Since
one expect that the motion of particles under the Galilean transformation
is isotropic, the $t^{-d/4 }$ decay law of the isotropic diffusion
is also expected in uniformly driven systems. Furthermore, for the
single species reaction $kA \to 0$, the anisotropy of diffusion is
no effect on the kinetics of HC particles \cite{kA}. Hence the
result of Janowsky is rather surprising. For the uniformly driven $A+B \to 0$
reaction without HC exclusion, the uniform drift also
does not change the kinetics \cite{ABcol}. Hence the interplay of the uniform
drift and HC constraint is crucial for the anomalous $t^{-1/3}$ decay law.

As an attempt to understand the origin of the anomalous decay law
$t^{-1/3}$, Ispolatov et al suggested a scaling argument provided
the motion of particles in a single species domain follows the
Burgers equation \cite{HC2}. From the drift velocity of a single
interface between $A$ and $B$ domains, they found the rate of the change
of the domain length $\ell$, defined as the distance between the
first particles of adjacent opposite species domains, scales to the
density $\rho$ as $d \ell /dt \sim \rho$. For random initial
conditions of $\rho_A (0) =\rho_B (0)$, $\rho$ scales as $\rho \sim
1/\sqrt{\ell}$ \cite{kang} so $\ell$ scales as $\ell \sim t^{2/3}$
instead of $t^{1/2}$ of the isotropic case. Hence, the anomalous
scaling of the domain length $\ell$ may result in the anomalous
density decay. In $d$ dimensions, $\ell$ scales as $\ell \sim
t^{(5-d)/6}$. So $\rho$ decays as $\rho \sim t^{-(d+1)/6}$ for $d
\leq 2$, $t^{-d/4}$ for $2 < d \le d_c(=4)$ \cite{HC2}. In $d>2$
dimensions, the drift is irrelevant.
In addition to the domain length $\ell$, the inter-particle
distance($\ell_{AA}$) and the distance between two adjacent
particles of opposite species($\ell_{AB}$)  also characterize the
spacial organization of particles \cite{length}. Janowsky
numerically showed that $\ell_{AA}$ and $\ell_{AB}$ scale as
$\ell_{AA} \sim t^{1/3}$ and $\ell_{AB} \sim t^{3/8}$ respectively \cite{HC3}.
Intriguingly, the scaling of $\ell_{AB}$ is not changed by the
drift. Using the scaling of $\ell_{AB}$, it was suggested that the
domain length $\ell$ scales as $t^{7/12}$ \cite{HC3}.
The extensive simulation results of Ref. \cite{HC4} supported the result of
Janowsky, but it conflicts with the prediction of Ref. \cite{HC2}.

Recently there have been some attempts to describe systems of HC particles
in field-theoretic formalism \cite{rgabhc,kArg,cphc}.
S.-C. Park et al systematically derived the $t^{-1/3}$ decay law of $\rho$.
The scaling of $\rho$ is well understood, while the scaling behaviors of various
lengths are not convincingly conclusive. Since the asymptotic scaling regime is extremely
slowly approached due to strong corrections \cite{HC2,HC4}, it is hard to observed the true
scaling behaviors directly via simulations.
Hence it is desired to study another model which exhibits
the same scaling behavior, but reaches the true scaling regime within moderate simulation time.
On the other hand, HC interaction can be generalized by allowing multiple occupation with
a certain attractive interaction between particles on the same site.
The on-site attractive interaction would give more general understanding about the scaling
behavior HC particles exhibit.

For this aim, in this paper, we introduce a one dimensional two species bosonic reaction model
with mass-dependent fragmentation.
In this model, the HC constraint is replaced by mass-dependent fragmentation of
bosonic particles. For simplicity, we consider the driven case to the right direction only.
If the randomly selected site $i$ is occupied by $n_i$ identical particles or mass $n_i$,
the mass $m= n^{\lambda}_i$ moves to the right nearest neighbor with unit rate.
The fragmentation of mass $m$ corresponds to the moves of randomly selected $m$ particles
among $n_i$ particles. When the opposite species of mass $n_{i+1}$ occupies the site $i+1$,
the reaction $n_{i+1} \to n_{i+1}-m$ occurs instantaneously. For $m > n_{i+1}$, $m-n_{i+1}$
mass occupies the site $i+1$.
For $\lambda=1$, the whole mass $n_i$
moves to the right so $\lambda=1$ corresponds to the uniformly driven bosonic model.
For $\lambda < 1$, the fragmented mass is less than $n_i$, which can be regarded as
the consequence of the attractive interaction between particles occupying the same site.
Our model generalize the HC interaction and enables to
investigate how the kinetics of the reaction depends on the interaction strength.

\begin{figure}
\includegraphics[scale=0.6]{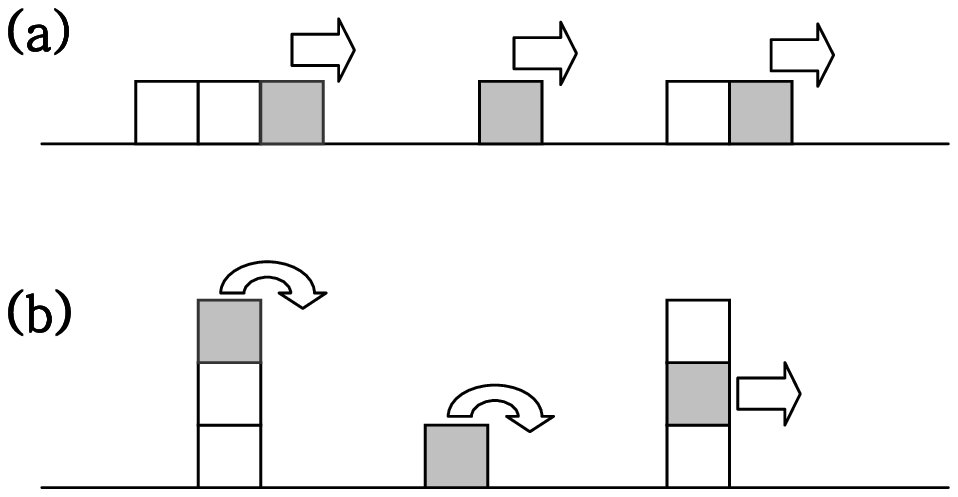}
\caption{\label{motion}Motions of particles driven to the right.
(a) Motions of HC particles. Only boundary particles (gray boxes) can hop to the right.
(b) Motions of bosonic particles with $\lambda=0$. A randomly selected particle (gray box)
hops to the right.
}
\end{figure}
Especially, for $\lambda=0$, only one particle or unit mass hops to the right with
unit rate. For HC particles, since only the right boundary particle of a train of
particles can hop with unit rate, the $\lambda=0$ case corresponds to HC interaction
(Fig.~\ref{motion}).
In some bosonic lattice gas model in which at each site, only the top particle on
the pile of particles hops to the bottom of the pile of the nearest neighboring site.
In this model, the order of particles is conserved so we call this model the ordered
bosonic lattice gas (BLG).
It was shown for the drift case that the motion of particles follows the Burgers equation as
in HC particle systems \cite{Kutner}. Since particles on a site are identical,
the ordered BLG describes the motion of particles inside single species domains
of the $\lambda=0$ case in our model.
For this reason, we regard HC interaction as the on-site attractive interaction between
bosonic particles in our model.
Hence the kinetics of $\lambda=0$ is expected to be the same as that of HC particles.
For $0<\lambda<1$, the number of hopping particles from a site ($m$) is larger than one,
but still smaller than $n_i$. Hence the interaction is weaker than HC constraint.
However, the interaction of $\lambda <0$ is stronger than HC interaction due to $m <1$.

We study the kinetics of the reaction $A+B \to 0$ with mass-dependent fragmentation rate
$D(m)=m^{\lambda}$ for
random initial distributions of equal densities of $A$ and $B$. With uniform drift to the one
direction of both species, we find for several values of $\lambda$ that the scaling of $\rho$,
$\ell_{AA}$ and $\ell_{AB}$ coincide with the results of Ref. \cite{HC3,HC4}.
However the domain length $\ell$ scales as
$\ell \sim t^{2/3}$ of Ref. \cite{HC2} rather than $t^{7/12}$ of Ref. \cite{HC3}.
Interestingly, the observed scaling
behaviors of density and lengths seem to be independent of $\lambda$ as long as $\lambda <1$.
In next section, we introduce our model in detail and present simulation results of several
values of $\lambda$. In Sec. III, we discuss the collective motion of single species domain
to understand the $\lambda$-independent kinetics of the reaction.
Finally, we conclude with summary in Sec. IV.

\section{Model and Simulation results}
On one dimensional lattice of size $L$ with periodic condition,
we consider a bosonic $A+B \to 0$ reaction with mass-dependent fragmentation
uniformly driven to the right.
First, randomly select a site. If the selected site $i$ is
occupied by mass $n_i$, $m(=n^{\lambda}_i )$ mass hops to the right with unit probability.
When the opposite species of mass $n_{i+1}$ occupies the site $i+1$,
the reaction $n_{i+1} \to n_{i+1}-m$ occurs instantaneously. For $m > n_{i+1}$, $m-n_{i+1}$
mass occupies the site $i+1$.
The color of mass at the site $i+1$ can be changed when $m > n_{i+1}$.
If the same species occupies the site $i+1$, the coagulation of $n_{i+1} \to n_{i+1}+m$
takes place.

We use the following algorithm for determining
the mass fragmented from a site $i$.
For $\lambda >0$, the mass fragmented from site $i$ ($n^{\lambda}_i$)
is not an integer number. In that case, integer $[n^{\lambda}_i ]$ mass hops and then
unit mass hops with probability $n^{\lambda}_i - [n^{\lambda}_i ]$.
$[x]$ denotes the integer number not greater than the real number $x$.
For $\lambda \leq 0$, only unit mass
attempts to hop with the probability $n^{\lambda}_i$.
With random initial distributions of $\rho_A (0) = \rho_B (0)=1/2$,
we perform Monte Carlo simulations on a chain of size $L = 1 \times 10^7$ for $\lambda = -1$
and $L=3\times 10^6$ for other values of $\lambda$ up to $t = 10^7$. We average densities and
various lengths over from $50$ to $150$ independent runs.

\begin{figure}
\includegraphics[scale=0.45]{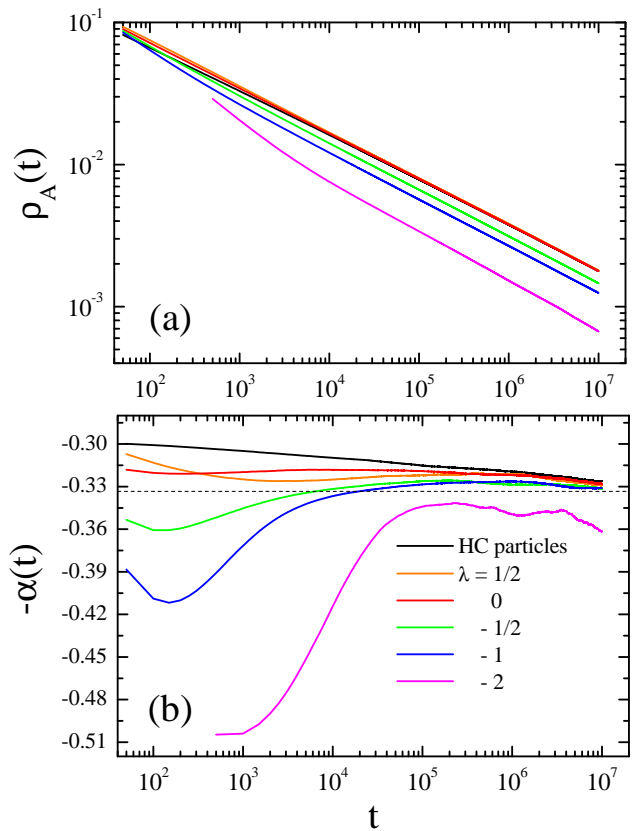}
\caption{\label{dens} (Color online)
(a) Double logarithmic plot of $\rho_A$. (b) The effective exponent $\alpha(t)$ and the
dashed line is $\alpha=1/3$. The color of each line is same in both panel.
}
\end{figure}

In Fig. \ref{dens}, we plot the density of $A$ species ($\rho_A (t)$) and its effective exponent defined as
$- \alpha(t) = \log[\rho_A (t)/\rho_A (t/b)]/\log b$ with $b=5$ for
$\lambda = 1/2, 0,-1/2,-1,-2$ respectively. For comparison with HC particles, we also plot the
results of HC particles.
With $\rho_A (0)= \rho_B (0)$,
the density $\rho_A (t)$ decays as $\rho_A \sim t^{-\alpha}$, and
decays faster than that of HC particles in all stages for $\lambda <1$.
Especially in early stage, the density decay becomes faster as $\lambda$ decreases for $\lambda<0$,
which is reflected by bumps in $\alpha (t)$. After the early stage,
$\rho_A (t)$ slowly gets into the asymptotic scaling regime of $\alpha = 1/3$.
From the scaling plot of $t^{\alpha}\rho(t)$ against $t$, we estimate $\alpha =0.329(2)$,
$0.331(3)$, $0.333(2)$ and $0.331(2)$ for $\lambda=1/2$, $0$, $-1/2$ and $-1$ respectively
which agree well with previous studies \cite{HC1,HC2,HC3,HC4}.
For HC particles, we estimate $\alpha=0.327(3)$. Hence our model approaches to
the scaling regime faster than HC particle system does.
For $\lambda=-2$, we estimate $\alpha=0.37(1)$. It indicates that
the asymptotic scaling region is not reached yet.
The physical argument for the slow approach to the asymptotic scaling of the
$\lambda=-2$ case is discussed in the following paragraph.

\begin{figure}
\includegraphics[scale=0.45]{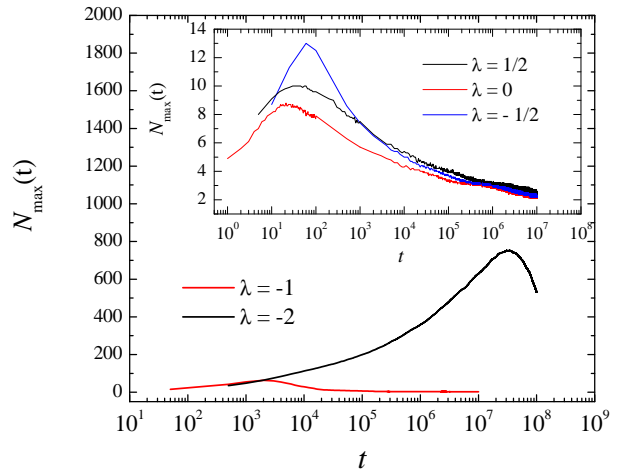}
\caption{\label{nmax} (Color online)
Semi-logarithmic plot of $N_{max}$ versus $t$. The main plot shows $N_{max}$ of $\lambda=-1$,
$-2$. In the inset, $N_{max}$ of $\lambda= 1/2$, $0$, $-1/2$ are plotted.
}
\end{figure}

To understand the early decay, we measure the maximum number of particles occupying
a single site at time $t$, $N_{max}(t)$. By definition, we have $n_i (t) \leq N_{max}(t)$.
Fig. ~\ref{nmax} shows the plot of $N_{max}(t)$ versus $t$ for several $\lambda$ values.
In early time, like particles pile up on sites due to multiple occupation. Hence
$N_{max}$ increases to a certain maximum value, and after then decreases monotonously.
The maximum of $N_{max}$ and the time it takes to reach the value tend to diverge
as $\lambda \to - \infty$.
Hence the time it takes to reach the
asymptotic scaling regime of $N_{max} \sim O(1)$ is also diverge as $\lambda \to - \infty$.
For example, the maximum of $N_{max}$ for $\lambda=-2$ is reached at $t \sim 3 \times 10^7$.
It means that the asymptotic scaling regime of $N_{max} \sim O(1)$ is very slowly approached
after $t \gg 3 \times 10^7$. Hence the system already feels the finite size before $t=10^7$,
which results in the larger estimate of $\alpha$.
For this reason, we perform simulations only for $\lambda > -2$.

Since the drift velocity of a particles on a site with large mass
is smaller than that on a site with small mass,
the annihilations of adjacent $AB$ pairs are accelerated.
For $\lambda \leq 0$,
it can be easily understood by considering the relative drift velocity of adjacent opposite
species. When the masses of two adjacent sites $i$ and $j$ ($j<i$) occupied by opposite species
are $n_i$ and $n_{j}$ respectively, the relative velocity of a particle at site $i$ to $j$ is
$v_{ij} = n_{i}^{\lambda}- n_{j}^{\lambda} = (1-(n_i /n_j )^{-\lambda})/n_{i}^{-\lambda}$.
For $n_i > n_j$, $v_{ij}$ is negative so the reaction is accelerated. On the other hand, for
$n_i < n_j$, a particle on the site $j$ is hard to react with a particle
on the site $i$ due to $v_{ij} >0$.
For $v_{ij}=0$ of the $n_i = n_j$ case, diffusive motions lead to the reaction as in HC particle
system.
As a result, the reactions are much more enhanced between domain boundaries of $v_{ij} <0$,
and also more accelerated as $\lambda \to - \infty$ in early stage.
Although $N_{max}$ decreases as the scaling regime is approached, the enhanced reactions by the
mass difference should be still dominant due to $N_{max} >1$. For this reason, the densities
of $\lambda <1$ decay faster than that of HC particles in the whole stage.

Next, we present simulation results of various lengths characterizing the spatial organization
of masses. The average length of a single species domains($\ell$) is defined as the
distance between the first mass of two adjacent opposite species domains.
The length $\ell_{AA}$ and $\ell_{AB}$ are defined as the inter-mass
distance between two adjacent masses of the same species and of opposite species
respectively \cite{length}. The lengths algebraically increase in time $t$
for $\rho_A (0) = \rho_B (0)$ as

\begin{equation}
\label{len}
\ell_{AA} \sim t^{1/z_{AA}},\;\; \ell_{AB} \sim t^{1/z_{AB}}, \;\; \ell \sim t^{1/z}.
\end{equation}

\begin{figure}
\includegraphics[scale=0.42]{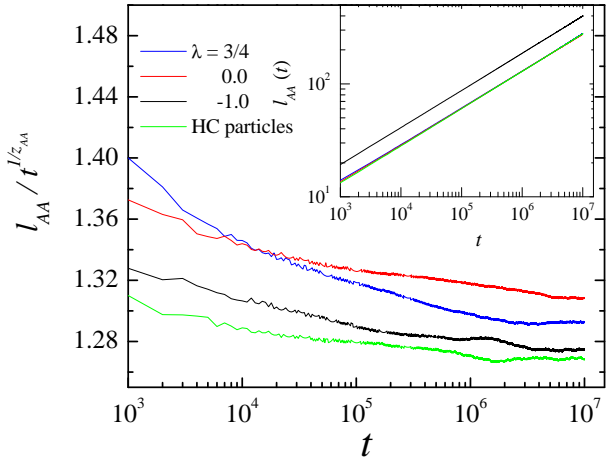}
\caption{\label{laa} (Color online)
The scaling plot of $\ell_{AA}$ against $t$.
The main plot shows the scaling plot of $\ell_{AA} t^{-1/z_{AA}}$. We use $1/z_{AA}=1/3$
for $\lambda = 3/4$, HC particles and $0.332$ for $\lambda=0,-1$.
We add $-0.62$ to the scaled value of $\lambda=-1$ and $-0.02$ to that of HC particles
for the better presentation.
The inset shows the double
logarithmic plot of $\ell_{AA}$ against $t$.
}
\end{figure}

\begin{figure}
\includegraphics[scale=0.42]{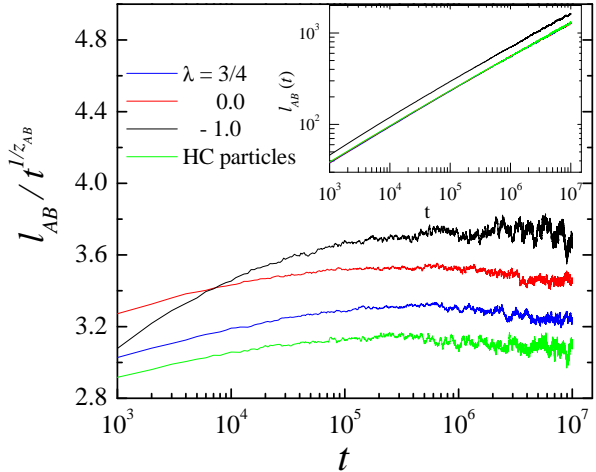}
\caption{\label{lab} (Color online)
The scaling plot of $\ell_{AB}$ against $t$.
The main plot shows the scaling plot of $\ell_{AB} t^{-1/z_{AB}}$. We use $1/z_{AB}=3/8$
for $\lambda = 3/4$, $0$, HC particles and $0.37$ for $\lambda=-1$.
We add a constant $0.2$, $0.4$ and $0.5$ to the scaled value of $\lambda =3/4$, $0$, and $-1$
respectively to avoid overlap.
The inset shows the double
logarithmic plot of $\ell_{AB}$ against $t$
}
\end{figure}

In the measurement of lengths, the periodic boundary condition is not used for better statistics.
Fig.~\ref{laa} shows the scaling plots of $\ell_{AA}t^{-1/z_{AA}}$ for $\lambda = 3/4,0,-1$.
We obtain the best scaling plots with $1/z_{AA}=1/3$ for $\lambda=3/4$ and
$0.332$ for $\lambda=0,-1$ respectively.
For HC particle system, the best scaling plot is obtained with $1/z_{AA}=1/3$ as expected.
For the length $\ell_{AB}$ (Fig.~\ref{lab}), we obtain the best scaling plot
of $\ell_{AB}t^{-1/z_{AB}}$ with
$1/z_{AB}=3/8$ for $\lambda=3/4$, $0$ and $0.37$ for $\lambda=-1$ respectively.
For HC particles, $1/z_{AB}=3/8$ gives a nice scaling plot as expected.
The simulation results of $\ell_{AA}$ and $\ell_{AB}$ agree very well with
the prediction $1/z_{AA}=1/3$ and $1/z_{AB}=3/8$ of previous studies \cite{HC3,HC4}.

\begin{figure}
\includegraphics[scale=0.42]{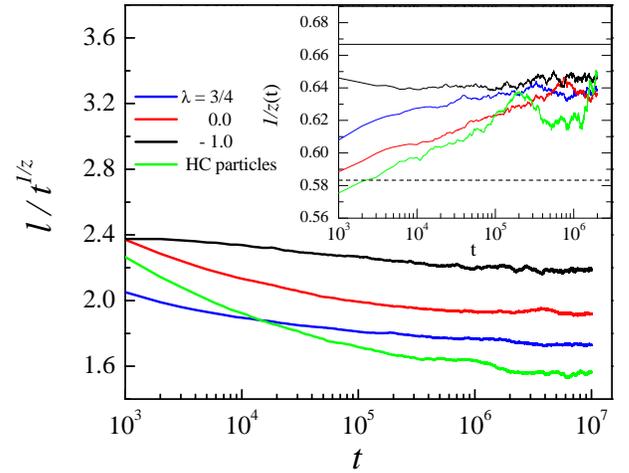}
\caption{\label{l} (Color online)
The scaling plot of $\ell$ against $t$.
The main plot shows the scaling plot of $\ell t^{-1/z}$. We use $1/z = 0.65$
for $\lambda = 3/4$, $-1$, HC particles and $0.64$ for $\lambda= 0$. The inset shows the
effective exponent $1/z(t)$. The solid and dashed horizontal lines correspond to $1/z =2/3$ and
$7/12$ respectively.
}
\end{figure}

For the domain length $\ell$, $1/z=2/3$ is predicted by Ref. \cite{HC2},
while $1/z=7/12$ is predicted and are numerically supported by
Ref. \cite{HC3,HC4}. Our simulation results support $1/z=2/3$ rather than $7/12$.
Fig.~\ref{l} shows the scaling plots of $\ell t^{-1/z}$ for $\lambda=3/4,0,-1$ and HC particles.
We obtain the best scaling plots with $1/z=0.65$ for $\lambda=3/4$, $-1$, HC particles and
$0.64$ for $\lambda=0$ respectively.
These values are close to the prediction $1/z=2/3$ of Ref. \cite{HC2}.
The inset shows the effective exponent $1/z(t)$ of $\ell (t)$ defined as
$\ln(\ell(bt)/\ell(t))/\ln b$ with $b=5$.
The effective exponent shows that our model reaches asymptotic region
faster than HC particle system does.
On the other hand, $1/z=7/12$ indicates the violation of the relation $\rho \sim 1/\sqrt \ell$
in the drift case, while $1/z=2/3$ satisfies the relation.
We check the relation of $\rho \sim 1/\sqrt \ell$ by investigating the scaling plot of
$\rho \ell^x$. We obtain the best scaling plot with $x=0.51(2)$ for several $\lambda$ and
HC particle system (not shown), which numerically confirm the relation $\rho \sim 1/\sqrt \ell$
in this uniformly driven case.
With $\alpha = 1/3$, we also expect $1/z=2/3$ from the relation $\rho \sim 1/\sqrt \ell$.
Therefore our numerical results agree well with the prediction of $1/z=2/3$ \cite{HC2}.

Our simulation results indicate that the kinetics of the reaction with mass-dependent
fragmentation does not depend on the exponent $\lambda (<1)$.
In particle version, $m=n^{\lambda}_i$ particles hop to the right at the
same time in our model, while only the boundary particle can hop in HC particle system.
The fragmented mass $m$ can
be regarded as $m$ particles selected in various ways such as random selection
or ordered selection from the top of the pile with $n_i$ particles.
Hence no direct mapping between our model and HC system is available.
For the ordered BLG model of Ref. \cite{Kutner},
it was shown that the dynamics of the density
described by Burgers equation and the dispersion of the center of mass (CM) scales as $t^{4/3}$
which is the characteristic of asymmetric exclusion process (ASEP) \cite{Gunter}.
For the same kinetics of our model as that of HC system, collective motions of single species
domain should share the feature of ASEP.
However, to the best of our knowledge, there are no such studies on bosonic model
with the present kind of mass-dependent fragmentation.
In next section, we discuss the collective diffusion of driven
single species domain with the mass-dependent fragmentation.

\section{The collective motion of a single domain}
For simplicity, we consider the bias to the right only. Hence a particle hops to the
right with unit probability. With initial density of $\rho$, particles are randomly
distributed on a ring of size $L$. When randomly selected site $i$ is occupied by
$n_i$ particles, $m=n^{\lambda}_i$ particles randomly selected and moves to the right
at the same time. When the site $i+1$ has $n_{i+1}$ particles, the mass coagulation of
$n_{i+1} \to n_{i+1}+m$ takes place instantaneously.
For $\lambda=1$, all $n_i$ particles move and only coagulation occurs without fragmentation.

We measure the dispersion of the displacement of the center of mass ($\sigma^{2}_{CM}$)
defined as $\sigma^{2}_{CM}=\overline{X^{2}} _{CM}-\overline{X}_{CM} ^{2}$.
$X_{CM}$ is the displacement of the center of
mass defined as $X_{CM}=\sum^N_i x_i /N$,
where $N$ is the total number of particles and $x_i$ is the displacement of $i$th particle
from its initial position. The dispersion $\sigma^2 _{CM}$ is 0 for driven bosonic particles and
scales as $t^{4/3}$ for driven HC particles \cite{Gunter}.
We perform Monte Carlo simulation for various values of $\lambda$ from $1/2$ to $-1$ on
chains of size $L=10^4$ with periodic boundary condition.
We average $\sigma^{2}_{CM}$ over $3\times 10^4$ independent runs
up to $t=10^4$ time steps.
In simulations, we use the exact $\rho$ and set $\rho=0.2$ for $\lambda=1/2$, $0.5$
for $\lambda=0$ and $0.01$ for $\lambda=-1$ respectively.

\begin{figure}
\includegraphics[scale=0.42]{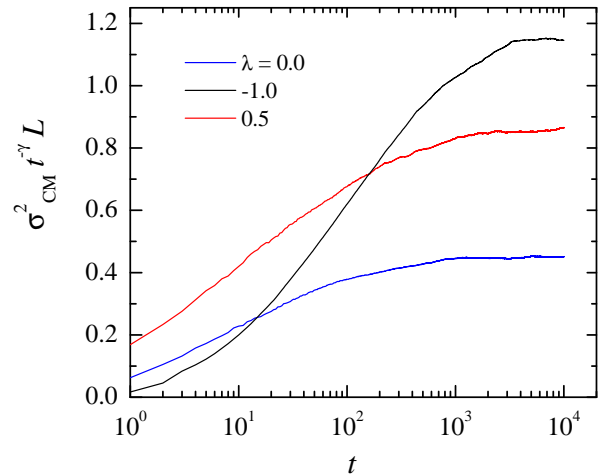}
\caption{\label{cm} (Color online)
The scaling plot of $\sigma^{2}_{CM} t^{-\gamma}L$ against $t$. We use $\gamma = 1.38$
for $\lambda = 0.5$, and $1.37$ for both $\lambda=-1$ and $0$ with system size $L=10^4$.
We multiply the line of $\lambda=-1$ by $0.04$ to draw the line near the others.
}
\end{figure}

Fig.~\ref{cm} shows the plot of $\sigma^{2}_{CM} t^{-\gamma} L$
against $t$. $\sigma^{2}_{CM}$ crossovers to $t^\gamma$ scaling regime.
We obtain the
best scaling plot with $\gamma = 1.38(1)$ for
$\lambda=1/2$, and $1.37(1)$ for $\lambda=-1$ and $0$ respectively which is slightly larger than
$\gamma=4/3$ of HC particles.
In the ordered BLG model of Ref. \cite{Kutner}, the velocity of CM is given as
$v_{CM}=1/(1+\rho)$ for the extreme drift to the one direction.
For $\lambda=0$ with $\rho=0.5$, we estimate $v_{CM}=0.669(1)$ which also
agree well with the prediction of Ref. \cite{Kutner}. For $\lambda=0$ with other densities,
we also confirm the prediction. Hence we are convinced that
the $\lambda=0$ case shares the same feature of
the ordered BLG model in spite of rather large numerical values
of $\gamma$.
On the other hand, for $\lambda \neq 0$, there are no theoretical predictions for
$v_{CM}$ and $\sigma^{2}_{CM}$. However the exponent $\gamma$ indicates that the collective
motions of $\lambda<1$ close to that of HC particles rather than that of bosonic particles.


We conclude that the collective driven motions of masses with mass-dependent fragmentation
share the feature of HC particles as long as $\lambda <1$.
It results in $\lambda$-independent kinetics of $A+B \to 0$ reaction with the mass-dependent
fragmentation.

\section{Summary}

In summary, we investigate the kinetics of uniformly driven
$A+B \to 0$ reaction with mass-dependent fragmentation.
In this model, the fragmented mass $m$ of a site with mass $n_i$ is given
as $m=n^{\lambda} _i$, and it is driven to the one direction. When opposite species mass
occupies the same site, mass reaction takes place instantaneously.
Since the fragmented mass $m$ of $\lambda<1$ is less than the whole mass $n_i$ of a site,
the exponent $\lambda$ controls the strength of the attractive interaction
between bosonic particles occupying the same site.
The $\lambda=0$ corresponds to the uniformly driven hard-core (HC) particles system because only
one particle hops with unit rate. Compared with $\lambda=0$,
the attractive interaction of $\lambda >0$ is weaker than HC interaction,
while the interaction of $\lambda<0$ is stronger.
Our model generalize the HC interaction and enables to
investigate how the kinetics of the reaction depends on the interaction strength.
Also we expect that our model gets into the asymptotic scaling regime
faster than HC particle system does,
since HC constraint is somewhat relaxed in our model by allowing any particle on a site to
move.

We perform Monte Carlo simulations in one dimension for several values of $\lambda$,
and measure the critical exponents of density and various lengths which exhibit power-law
scaling for $\rho_A (0)=\rho_B (0)$.
The scaling behaviors of density and lengths except the domain length $\ell$ are the
same as those of HC particles of Ref. \cite{HC1,HC3,HC4}. The scaling behavior of $\ell$
are close to $t^{2/3}$ of Ref. \cite{HC2} rather than $t^{7/12}$ of Ref. \cite{HC3}.
Interestingly, the kinetics of the reaction is independent of $\lambda$ as long as $\lambda<1$.
To understand $\lambda$-independent behavior, we investigate the collective driven
motions of single domains. The displacement of the center of mass
($\sigma^2 _{CM}$) scales as $t^{\gamma}$ with $\gamma=4/3$ for driven HC particles \cite{Gunter}
and $\sigma^2 _{CM}$ is zero for bosonic particles.
For various $\lambda$ values, we find that $\gamma$ is slightly larger than but close to $4/3$.
The collective motion of a single domain shares the feature of HC particles.
Furthermore the motion is independent of $\lambda$,
which results in $\lambda$-independent kinetics of the reaction.
However for the quantitative understanding of the collective motion, another analytical
study is desired.

This work is supported by Grant No. R01-2004-000-10148-0 from the
Basic Research Program of KOSEF.

\end{document}